\documentclass[12pt]{article}
\usepackage{epsfig}
\usepackage{rotating}

\begin{document}

\begin{center}
{\Large \bf Search for direct CP-violation in
$K^{\pm}\rightarrow\pi^\pm\pi^0\pi^0$ decays}
\end{center}

\begin{center}
{\Large The NA48/2 Collaboration}\\
\vspace{2mm}
J.R.~Batley, A.J.~Culling, G.~Kalmus, C.~Lazzeroni, D.J.~Munday,
M.W.~Slater,
S.A.~Wotton \\
{\em \small Cavendish Laboratory, University of Cambridge,
Cambridge, CB3 0HE,
U.K.$\,$\footnotemark[1]} \\[0.2cm]
R.~Arcidiacono, G.~Bocquet, N.~Cabibbo, A.~Ceccucci,
D.~Cundy$\,$\footnotemark[2], V.~Falaleev, M.~Fidecaro, L.~Gatignon,
A.~Gonidec, W.~Kubischta, A.~Norton, A.~Maier, M.~Patel,
A.~Peters\\
{\em \small CERN, CH-1211 Gen\`eve 23, Switzerland}
% ...
\\[0.5cm]
%
\rm
\end{center}

\setcounter{footnote}{0} \footnotetext[1]{Funded by the U.K.
Particle Physics and Astronomy Research Council}
\footnotetext[2]{Present address: Istituto di Cosmogeofisica del CNR
di Torino, I-10133 Torino, Italy} \footnotetext[3]{Present address:
CERN, CH-1211 Gen\`eve 23, Switzerland} \footnotetext[4]{Also at
Centre de Physique des Particules de Marseille, IN2P3-CNRS,
Universit\'e de la M\'editerran\'ee, Marseille, France}
\footnotetext[5] {Also Istituto di Fisica, Universit\`a di Modena,
I-41100  Modena, Italy} \footnotetext[6]{Istituto di Fisica,
Universit\`a di Urbino, I-61029  Urbino, Italy}
\footnotetext[7]{Funded by the German Federal Minister for Education
  and research under contract 05HK1UM1/1}
\footnotetext[8]{Permanent address: SLAC, Stanford University, Menlo
  Park, CA 94025, U.S.A.}
\footnotetext[9]{Present address: UCLA,  Los Angeles, CA 90024,
U.S.A.} \footnotetext[10]{Funded by the German Federal Minister for
Research and Technology (BMBF) under contract 056SI74}
\footnotetext[11]{Funded by the Austrian Ministry for Traffic and
Research under the contract GZ 616.360/2-IV GZ 616.363/2-VIII, and
by the Fonds f\"ur Wissenschaft und Forschung FWF Nr.~P08929-PHY}

\abstract

A search for direct CP-violation in
$K^{\pm}\rightarrow\pi^\pm\pi^0\pi^0$ decay based on 47.14
million events has been performed by the NA48/2 experiment at the
CERN SPS. The asymmetry in the Dalitz plot linear slopes
$A_g=\left(g^+-g^-\right)/\left(g^++g^-\right)$ is measured to be
$A_g=(1.8\pm2.6)\cdot10^{-4}$. The design of the experiment and the
method of analysis provide good control of instrumental
charge asymmetries in this measurement. The precision of the result
is limited by statistics and is almost one order of magnitude better
than that of previous measurements by other experiments.

\newpage

\section{Introduction}

Studies of direct CP-violation play an important role in
understanding the nature of weak interactions, and
provide the opportunity to search for
physics beyond the Standard Model (SM). More than three decades
passed since the discovery of CP-violation in the mixing of
neutral kaons~\cite{1}, until
direct CP-violation was definitively established in the neutral kaon
system by the measurement of a non-zero $\epsilon^\prime/\epsilon$
parameter~\cite{3},\cite{4}. More recently, direct CP violation
has been also detected in B meson decays~\cite{2}. In order to explore
possible non-SM enhancements to heavy-quark loops which are at the
core of direct CP-violating processes, different systems
must be studied. In kaons, besides
the  $\epsilon^\prime/\epsilon$ parameter in
$K_L\rightarrow\pi\pi$ decays, promising complementary
observables are the rates of GIM-suppressed rare kaon decays
proceeding through neutral currents, and an asymmetry between $K^+$ and
$K^-$ decays to three pions. The $K^\pm\rightarrow 3\pi$ matrix
element is usually parameterized by a polynomial expansion in two
Lorentz-invariant variables $u$ and $v$:
\begin{equation}
\left|M\left(u,v\right)\right|^2\propto1+gu+hu^2+kv^2+...,
\end{equation}
where $|h|, |k|\ll|g|$ are parameters, and
\begin{equation}
u=\frac{s_3-s_0}{m^2_\pi},\quad{} v=\frac{s_1-s_2}{m^2_\pi},
\end{equation}
where $m_\pi$ is the charged pion mass,
$s_i=\left(p_K-p_i\right)^2$, $s_0=\sum s_i/3$ $(i=1,2,3)$, $p_K$
and $p_i$ are kaon and $i$-th pion 4-momenta, respectively. The
index $i=3$ corresponds to the odd, i.e. charged pion (the other two
pions have the same charge). Any difference in the slope parameters
between $K^+$ and $K^-$ indicates the presence of direct CP
violation. SM predictions for the asymmetry in the linear slopes
$g^+$ for $K^+$ and $g^-$ for $K^-$:
\begin{equation}
A_g=\frac{g^+-g^-}{g^++g^-}
\end{equation}
vary from a few $10^{-6}$ to a few $10^{-5}$~\cite{5}. Several
experiments~\cite{6} have searched for such asymmetries with a
precision at the level of $10^{-3}$ for both
$K^\pm\rightarrow\pi^\pm\pi^+\pi^-$ and
$K^\pm\rightarrow\pi^\pm\pi^0\pi^0$ decay modes, obtaining no
evidence for direct CP violation. Existing
theoretical calculations involving processes beyond the SM~\cite{7}
do not exclude substantial enhancements of the asymmetry $A_g$
which could be observed in the present experiment.

The NA48/2 experiment at the CERN SPS was designed
to search for
direct CP-violation in the decays of charged kaons to three pions,
and collected data in 2003 and 2004. A result based on the analysis of
$4.714\cdot10^7$ $K^{\pm}\rightarrow\pi^\pm\pi^0\pi^0$ decays
accumulated during the 2003 run is presented in this paper. An
asymmetry result for $K^{\pm}\rightarrow\pi^\pm\pi^+\pi^-$ decays
recorded at the same time has recently been published~\cite{8}.

\section{Description of the Experiment}

In order to reach a high accuracy in the measurement of the charge
asymmetry parameter $A_g$, the highest possible level of charge
symmetry between $K^+$ and $K^-$  is a crucial requirement in the
choice of beam, experimental apparatus, strategy of data taking and
analysis. A novel beam line with two simultaneous charged beams of
opposite charges was designed and built in the high intensity hall
(ECN3) at the CERN SPS. The simultaneous recording of both $K^+$ and
$K^-$ decays and a frequent inversion of all magnetic field
polarities provide a high level of intrinsic cancellation of the main
possible systematic effects in the measurement of $A_g$.
\begin{figure}[t]
\begin{center}
\resizebox{.9\textwidth}{!}{\includegraphics{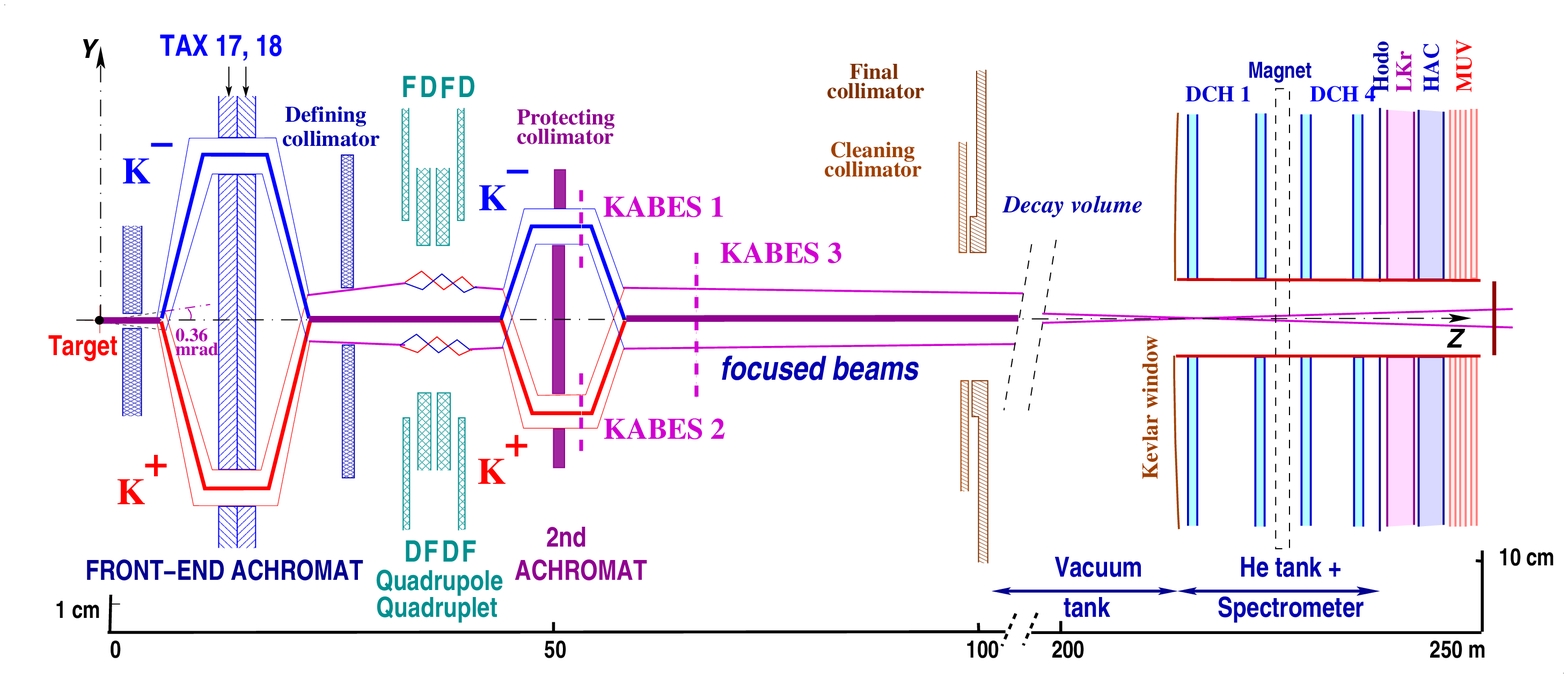}}
\end{center}
\vspace{-6mm} \caption{Schematic lateral view of the NA48/2 experiment.
Region 1 (from target to decay volume): beam line (TAX17,18: motorized
beam dump/collimators used to select the
momentum of the $K^+$ and $K^-$ beams; DFDF: focusing
quadrupoles; KABES1-3: beam spectrometer stations). Region 2: decay
volume and detector (DCH1-4: drift chambers; Hodo:
hodoscope; LKr: electromagnetic calorimeter; HAC: hadron
calorimeter; MUV: muon veto). The vertical scales for the
two regions are different.} \label{beamline}
\end{figure}

A schematic view of the beam line and detectors is presented in
Fig.~\ref{beamline}. A right-handed coordinate system is used to
describe the apparatus, with the $z$ axis along the beam line
direction, and the $y$ axis directed vertically up (the field in the
centre of the spectrometer magnet (see below) is alternatively parallel
or anti-parallel to the $y$ axis). The beam line and the set-up are
designed to be symmetric with respect to $x=0$. The charged particle
beams are produced by 400 GeV protons from the SPS impinging at zero
angle with respect to the $z$ axis on a 40 cm long beryllium target
with a diameter of 2 mm. The proton beam has an intensity of about
$7\times10^{11}$ protons per burst with a duty cycle of $\sim4.8$ s
/ 16.8 s. Charged particles with a central momentum of 60 GeV/$c$
and a momentum bite of $\pm$3 GeV/$c$ (FWHM) are selected
symmetrically by a first achromatic magnet system ('achromat') which
separates the two beams in the vertical plane and recombines them
again on the same axis. Both beams then pass through a series of
focusing quadrupoles and are again split and cleaned in a second
achromat. The second achromat together with the KABES
detector~\cite{11} also serves as a beam spectrometer (not used in
the present analysis). The $K^+/K^-$ flux ratio is about 1.8, and is
found to be stable during data taking.

Downstream of the second achromat both beams follow the same path and,
after passing the cleaning and final collimators, enter a 114 m long
evacuated decay region with a diameter of 1.92 m for the first
66 m, and 2.4 m for the rest. Here the two beams are collinear
and superimposed within one millimetre, until they reach a spectrometer
magnet where they are deflected horizontally in opposite directions
by an angle of ~2mrad.

The reconstruction of $K^{\pm}\rightarrow\pi^\pm\pi^0\pi^0$ decays
is based on the information from a magnetic spectrometer and a
liquid krypton calorimeter (LKr). The spectrometer is housed in a
tank filled with helium at atmospheric pressure separated from the
vacuum tank by a $kevlar$ window with a thickness of 0.0031 radiation
lengths ($X_0$). Surviving beam particles, as well as muons from
$\pi\rightarrow\mu\nu$ decays, traverse the centre of the
detectors inside an evacuated beam tube with a diameter of $\sim16$
cm. Two drift chambers (DCH1,2) are located upstream and two
(DCH3,4) downstream of a dipole magnet which deflects charged
particles in the horizontal plane with a
transverse momentum kick of 120 MeV/$c$. The DCHs
have an octagonal shape with an area of 4.5 m$^2$. Each chamber
has eight planes of sense wires, two horizontal, two vertical
and two along each of two orthogonal 45$^\circ$ directions.
The momentum resolution
of the magnetic spectrometer is $\sigma(p)/p=1.0\%\oplus0.044\%p$
($p$ in GeV/$c$). The magnetic spectrometer is followed by
a scintillator hodoscope consisting of two planes segmented into
horizontal and vertical strips and arranged in four quadrants.

The LKr calorimeter ~\cite{12} is an almost homogeneous ionization
chamber with an
active volume of  10 m$^3$ of liquid krypton, segmented
transversally into 13248 projective cells, 2$\times$2 cm$^2$ each, by
a system of Cu$-$Be ribbon electrodes, and with no longitudinal
segmentation. The calorimeter is 27 $X_0$ deep and has an energy
resolution $\sigma(E)/E=0.032/\sqrt{E}\oplus0.09/E\oplus0.0042$ ($E$
in GeV). The space resolution for a single electromagnetic shower
can be parameterized as $\sigma_x=\sigma_y=0.42/\sqrt{E}\oplus0.06$
cm for the transverse coordinates $x$ and $y$. The LKr is used to
reconstruct $\pi^0\rightarrow\gamma\gamma$ decays.

The NA48 apparatus also includes a hadronic calorimeter and a muon
detector. A description of these components can be found
elsewhere~\cite{12}.

Kaon decays are selected by a two-level trigger. The first level
requires a signal in at least one quadrant of both scintillator
hodoscope planes in coincidence with the presence of energy deposition
in the LKr consistent with at least two photons. At the second
level, a fast processor receiving the DCH information reconstructs
the charged particle momentum and calculates the missing mass
under the assumption that the particle is a $\pi^{\pm}$ originating
from the decay
of a 60 GeV/$c$ $K^\pm$ travelling along the nominal beam axis. The
requirement that the missing mass is not consistent with a $\pi^0$
mass rejects most $K^{\pm}\rightarrow\pi^\pm\pi^0$ events.
The typical rate of this trigger is $\sim$15,000 per
burst.

\section{Event selection and reconstruction}

After rejecting data affected by the malfunctioning of some detector
components, nearly 110 million events are selected for further
analysis, containing at least one charged
particle with momentum above 5 GeV/$c$ and at least four energy
clusters in the LKr, each consistent with a photon and above an energy
threshold of 3 GeV. In order to exclude events with overlapping
electromagnetic showers in the LKr, the distance
between any two photons in the LKr is required to be larger than
10 cm, and the distance between each photon and the impact point of the
charged particle on the LKr plane must exceed 15 cm.
Fiducial cuts on the distance
of each photon from the LKr edges and the central hole are also
applied in order to ensure full containment of the electromagnetic
showers. In order to symmetrize the geometrical acceptance to
particles of opposite charges, radial cuts are
applied around the average beam position, which is
continuously monitored for $K^+$ and $K^-$ separately using
the large sample of simultaneously recorded
$K^{\pm}\rightarrow\pi^{\pm}\pi^+\pi^-$ decays~\cite{8}.
In particular, to eliminate effects associated with the drop of
DCH efficiency near the central beam hole, the distance between
the charged pion and the average beam position on DCH1 (DCH4)
is required to be larger than 12.5 cm (16.5 cm).
Around 70 million events pass these requirements.

For each selected event, the $K^{\pm}\rightarrow\pi^\pm\pi^0\pi^0$
decay is reconstructed as follows.
Assuming that each pair $i,k$ of LKr clusters ($i,k=$1,2,3,4)
originates from a $\pi^0\rightarrow\gamma\gamma$ decay, the
distance $D_{ik}$ between the $\pi^0$ decay vertex position along
the $z$ axis and the front plane of the LKr is calculated:
\begin{equation}
D_{ik}=\frac{\sqrt{E_iE_k\left[\left(x_i-x_k\right)^2+\left(y_i-y_k\right)^2\right]}}{m_{\pi^0}},
\end{equation}
where $E_i$ and $E_k$ are the energies of the $i$-th and $k$-th
photon, respectively, $x_i$, $y_i$, $x_k$, $y_k$ are the coordinates
of their impact points on the LKr front plane, and $m_{\pi^0}$ is the
$\pi^0$ mass~\cite{13}.
\begin{figure}[t]
\begin{center}
\resizebox{.9\textwidth}{!}{\includegraphics{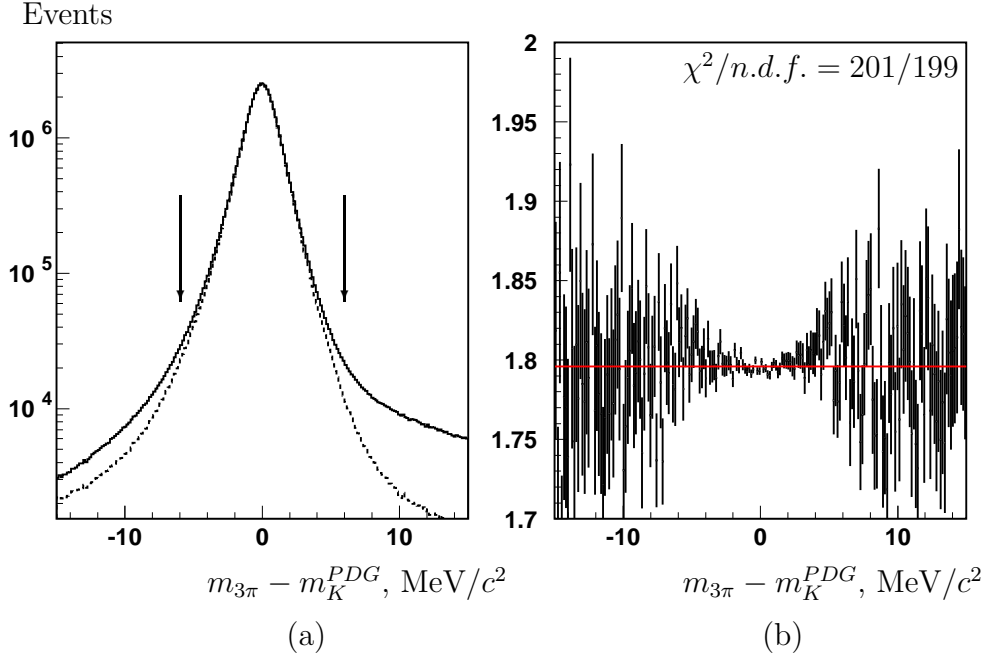}}
\put(-268,155){\vector(0,-1){40}} \put(-330,155){\vector(0,-1){40}}
\put(-140,5){$m_{3\pi}-m_K^{PDG}$, MeV/$c^2$}
\put(-320,5){$m_{3\pi}-m_K^{PDG}$, MeV/$c^2$} \put(-110,-15){(b)}
\put(-290,-15){(a)} \put(-390,220){Events}
\put(-140,200){$\chi^2/n.d.f.=201/199$}

\end{center}
\vspace{-6mm} \caption{(a) Difference between the invariant
$\pi^\pm\pi^0\pi^0$ mass and the PDG kaon mass~\cite{13}. The selected
interval is indicated by the arrows. The dashed
histogram shows the same distribution for events with no hit in the
muon detector.
(b) Ratio of invariant mass distributions for $\pi^+\pi^0\pi^0$ and
$\pi^-\pi^0\pi^0$ fitted with a constant function.}\label{mass}
\end{figure}

Among all photon pairs, the two with the smallest $D_{ik}$ difference are
selected as the best combination consistent with the hypothesis of
two $\pi^0$ mesons originating from $K^{\pm}\rightarrow\pi^\pm\pi^0\pi^0$
decay. The arithmetic average of the two $D_{ik}$ values is used
to define the distance of the reconstructed $K^{\pm}$ decay vertex
from the LKr (this choice gives a
$\pi^0\pi^0$ invariant mass equal to $2m_{\pi^0}$ at threshold,
corresponding to the best possible resolution).

Fig.~\ref{mass} (a) shows the invariant mass distribution for two
$\pi^0$s and a reconstructed charged particle track, assumed to be
$\pi^+$ or $\pi^-$. A clear signal from
$K^\pm\rightarrow\pi^\pm\pi^0\pi^0$ decays is seen, as expected,
with a mass resolution of $\sim$0.9 MeV/$c^2$. The tails originate
from wrong photon pairing (the fraction of these events is
0.23\%, as estimated by a Monte Carlo simulation) and
$\pi\rightarrow\mu\nu$ decays. No charge asymmetry is seen in this
distribution, as shown in Fig.~\ref{mass} (b).

Further event selection requires the $\pi^\pm\pi^0\pi^0$ invariant
mass to differ from the nominal $K^\pm$ mass~\cite{13} by less than
6 MeV/$c^2$, and the reconstructed kaon momentum to be between 54
and 66 GeV/$c$. In addition, the times of the track ($t^{\pm}$) and
of the photons ($t^{\gamma}$) must be consistent with a single event
within the experimental resolution: $\left|\langle t^\gamma\rangle
-t^\gamma_i\right|<5$ ns and $\left|\langle t^\gamma\rangle
-t^\pm\right|<20$ ns, where $\langle t^\gamma\rangle$ is the average
time of the four photons. These requirements are satisfied by
47.14$\times 10^6$ events.

The presence of magnetic fields in the spectrometer and to a lesser
extent in the beam line (achromats, focusing quadrupoles, etc.)
introduces a small but unavoidable charge asymmetry in the apparatus
acceptance. In order to effectively eliminate differences between
the $K^+$ and $K^-$ beams, the polarities of all magnets in the beam
line ('achromat polarities') are reversed weekly during data taking,
while the polarity of the spectrometer magnet is reversed daily. All
magnet currents are carefully monitored and kept at their nominal
value at the level of 10$^{-4}$. A set of data which contains at
least two periods with different achromat polarities is called a
'super-sample' and is treated in the analysis as an independent,
self-consistent data unit. During the 2003 running period three
super-samples were collected (see Table~\ref{t1})\footnote{The
super-sample (SS) notation used in this paper differs from that
of~\cite{8}: SS I and SS III correspond to SS 0 and SS 3
of~\cite{8}, respectively, while SS II contains data from both SS 1
and SS 2 corresponding to running conditions for which the
first-level trigger described in Section 2 was fully operational.}.
Each super-sample contains four $K^+\rightarrow\pi^+\pi^0\pi^0$ and
four $K^-\rightarrow\pi^-\pi^0\pi^0$ samples with different
combinations of the beam line and spectrometer magnet polarities.
\begin{table}[h]
\begin{center}
\begin{tabular}{|c|c|c|c|c|c|}
\hline
Super- & Sub- & \multicolumn{2}{|c|}{Achromat A+} & \multicolumn{2}{|c|}{Achromat A-} \\
\cline{3-6}
samples & samples & $K^+$ & $K^-$ & $K^+$ & $K^-$ \\
\hline \hline
I & 22 & 8.53 & 4.63 & 7.87 & 4.51 \\
\hline
II & 16 & 6.17 & 3.44 & 4.00 & 2.22 \\
\hline
III & 4 & 1.84 & 1.02 & 1.87 & 1.04 \\
\hline \hline

Total & 42 & \multicolumn{4}{|c|}{47.14} \\
\hline
\end{tabular}
\end{center}
\caption{Number of selected $K^+\rightarrow\pi^+\pi^0\pi^0$ and
$K^-\rightarrow\pi^-\pi^0\pi^0$ decays (in millions). A sub-sample
is a set of events taken in a run with a given configuration of
beam and spectrometer magnet polarities.} \label{t1}
\end{table}

\section{Asymmetry measurement method}
\begin{figure}[t]
\begin{center}
\begin{tabular}{cc}
\resizebox{.45\textwidth}{!}{\includegraphics{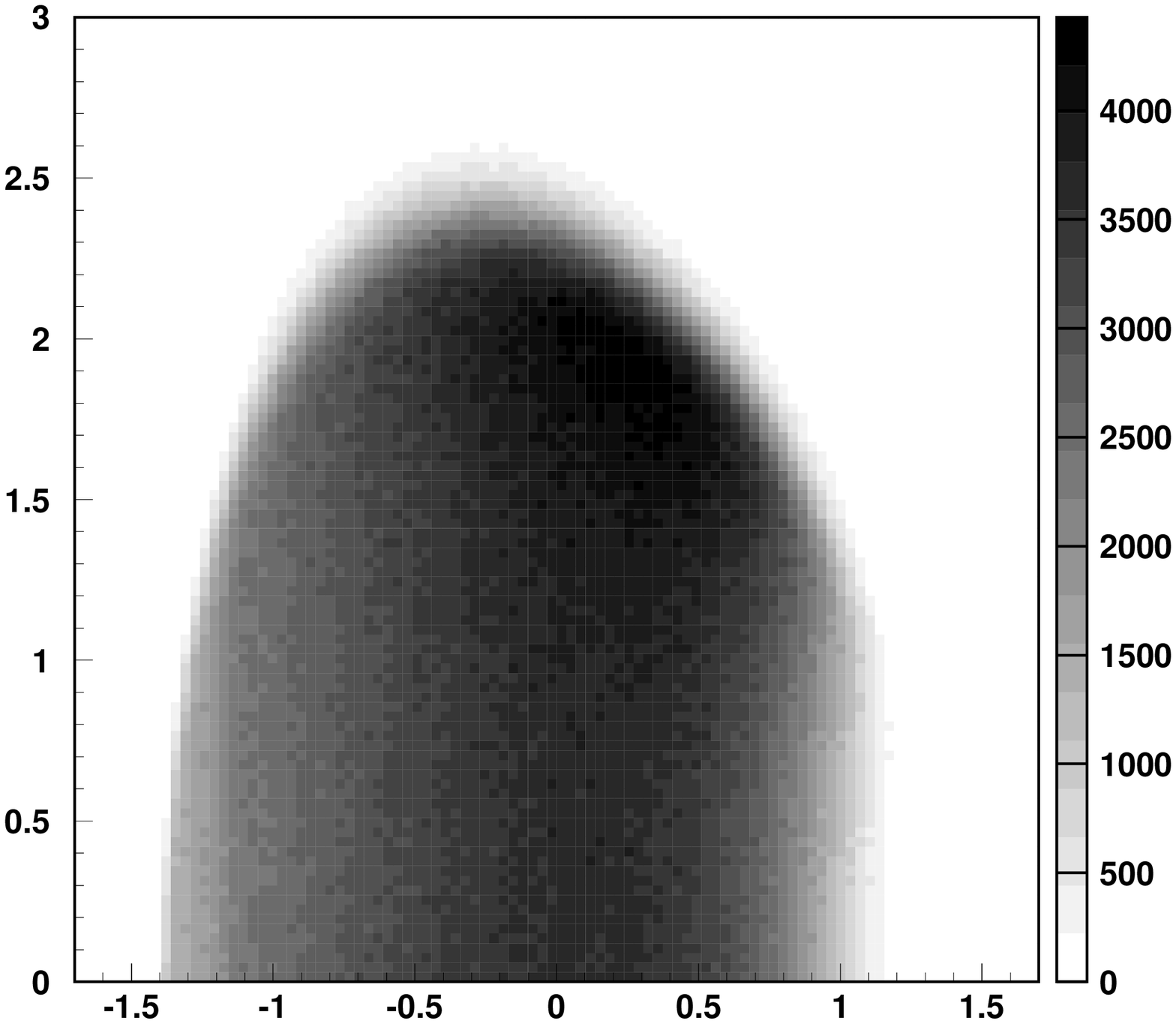}}
\put(-30,5){$u$} \put(-210,180){$|v|$} \hspace{0.5cm}
\resizebox{.45\textwidth}{!}{\includegraphics{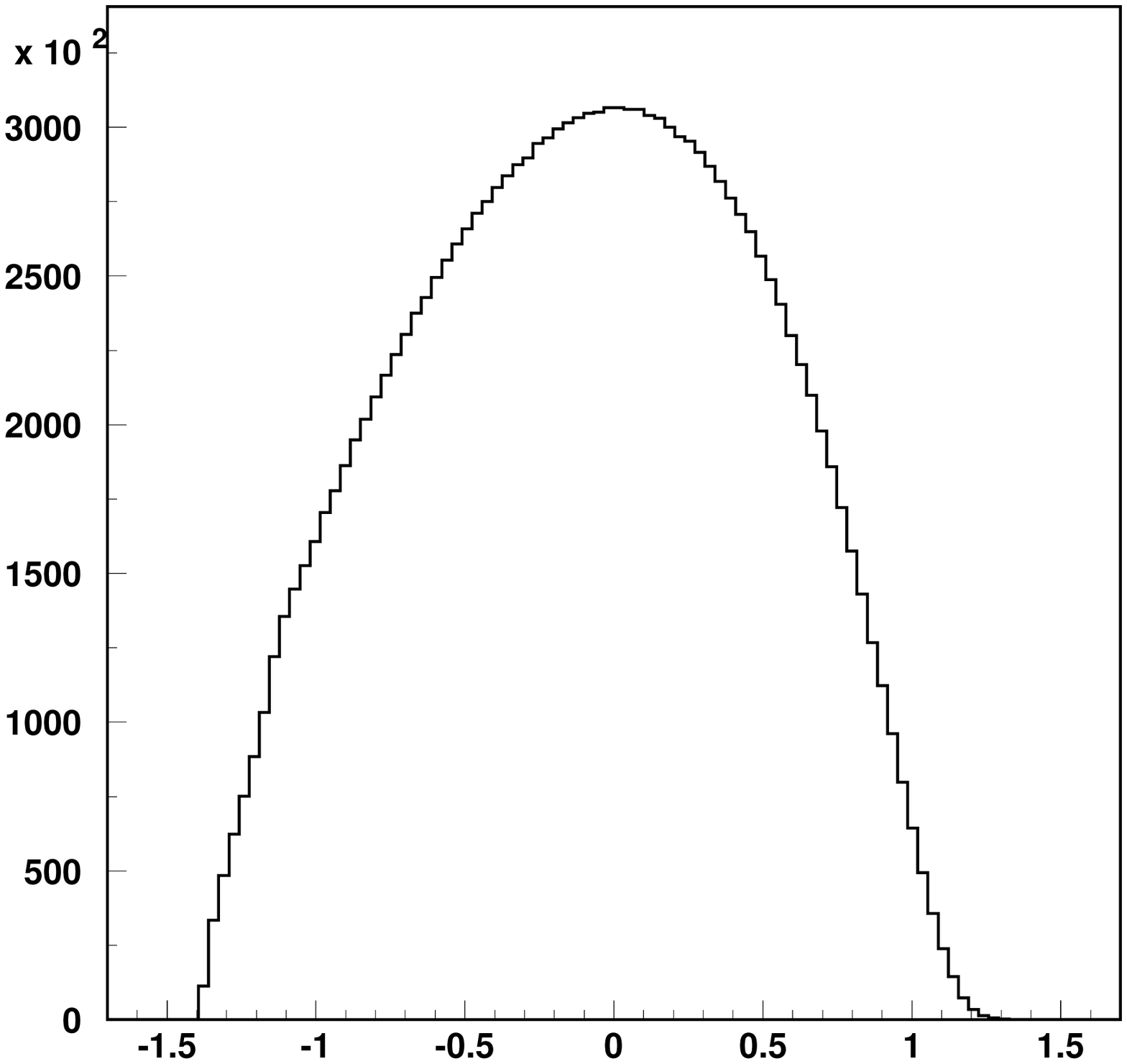}}
\put(-25,5){$u$} \put(-200,190){Events} \put(-110,-5){(a)}
\put(-330,-5){(b)}
\end{tabular}
\end{center}
\vspace{-6mm} \caption{(a) Dalitz plot folded around $v=0$; (b)
$u$-spectrum for the selected events.} \label{dalitz}
\end{figure}

The variable $u$ is defined as:
\begin{equation}
u=\frac{M^2_{00}-s_0}{m^2_\pi},
\end{equation}
where $s_0=(m^{2}_{K}+m^{2}_{\pi}+2m^{2}_{\pi^0})/3$,
$m_K$, $m_\pi$, $m_{\pi^0}$ are the kaon, charged
pion, neutral pion masses, respectively, and $M_{00}$ is the invariant
mass of two neutral pions evaluated using only information from the LKr
(a charge-blind detector).
The Dalitz plot of the selected events is shown
in Fig.~\ref{dalitz} (a). The events at large $u$ are affected most by
the detector acceptance because for these the charged pion trajectory
is very close to the beam pipe. Fig.~\ref{dalitz} (b) shows the $u$
spectrum (the projection of the Dalitz plot onto the $u$-axis) for
these events.

The asymmetry measurement is based on the comparison of the $u$ spectra
for $K^+$ and $K^-$ decays, $N^+(u)$ and $N^-(u)$, respectively.
For $K^{\pm}\rightarrow\pi^{\pm}\pi^0\pi^0$ decays the ratio of the
$u$ spectra $N^+(u)/N^-(u)$ is proportional to
\begin{equation}
\frac{N^+(u)}{N^-(u)}\propto\frac{1+(g+\Delta g)u+hu^2}{1+gu+hu^2},
\end{equation}
where $g=(0.638\pm 0.020)$ and $h=(0.051\pm0.013)$~\cite{13}. The
possible presence of a direct CP violating difference between the
linear slopes of $K^+$ and $K^-$, $\Delta g=g^+-g^-$,
can be extracted from a fit to this ratio. The measured
asymmetry is then given by $A_g=\Delta g/2g$.

In order to minimize the effect of beam and detector asymmetries
we use the ratio $R_4(u)$, defined as the product of four
$N^+(u)/N^-(u)$ ratios:
\begin{equation}
R_4(u)=R_{US}\cdot R_{UJ}\cdot R_{DS}\cdot
R_{DJ}=R\left(1+\frac{\Delta g\cdot u}{1+gu+hu^2}\right)^4,
\end{equation}
where the first subscript $U$ ($D$) denots a configuration of the beam magnet
polarities which corresponds to the positive beam traversing the
upper (lower) path in the achromats; the second subscript $S$ denotes the
spectrometer magnet polarities (opposite for the events in the
numerator and in the denominator of each ratio) deflecting the charged pions
to negative $x$ (towards the Sal\`{e}ve mountain,
given the topographical situation of the experiment in
relation to the mountains surrounding CERN) and $J$
corresponds to the deflection of the
charged pions in the opposite direction (towards the Jura mountain chain).
The spectra $N^+(u)$ and $N^-(u)$ for each of the four individual
ratios in (7) are obtained
from successive runs taken with the same
beam magnet polarities and with the $\pi^{\pm}$ deflected in the
same direction by the spectrometer magnet.
The parameter $\Delta g$ and the normalization $R$ are extracted
from a fit to the measured quadruple ratio $R_4(u)$
using the function in eq. (7). The measured
slope difference $\Delta g$ is insensitive to the normalization
parameter $R$, which reflects the ratio of $K^+$ and $K^-$ fluxes.

The quadruple ratio method complements the procedure of magnet
polarity reversal. It allows a three-fold cancellation of systematic
biases:

\begin{itemize}
\item beam line biases cancel between $K^+$ and $K^-$ samples in which
the beams follow the same path;
\item the effect of local non-uniformities of the detector
cancel between $K^+$ and $K^-$ samples in which charged pions
illuminate the same parts of the detectors;
\item as a consequence of using simultaneous $K^+$ and $K^-$ beams,
global, time dependent, instrumental charge asymmetries
cancel between $K^+$ and $K^-$ samples.
\end{itemize}

A reduction of possible systematic biases due to the presence of stray
permanent magnetic fields (Earth field, vacuum tank magnetization)
is achieved by the radial cuts around the average beam position (see
Section 3), which make the geometrical acceptance to charged pions
azimuthally symmetric. The only residual sensitivity to instrumental
charge asymmetries is associated with time variations
of any acceptance asymmetries occurring on a time scale shorter
than the magnetic field alternation period. However, their occurrence
would have been detected by a number of monitors recorded throughout
data taking.

Thanks to the symmetrical $K^+$ and $K^-$ decay acceptances, the
measurement does not require a Monte Carlo acceptance calculation.
Nevertheless, a detailed GEANT-based~\cite{14} Monte
Carlo simulation has been developed as a tool to study the sensitivity
of the result to systematic effects.
The Monte Carlo simulation includes an accurate
mapping of the spectrometer magnetic field (as well
as of the small stray permanent fields), full detector geometry and material
description, run-by-run simulation of time variations of local DCH
inefficiencies, and time variations of the beam properties.

\section{Result}
\begin{figure}[t]
\begin{center}
\resizebox{.8\textwidth}{!}{\includegraphics{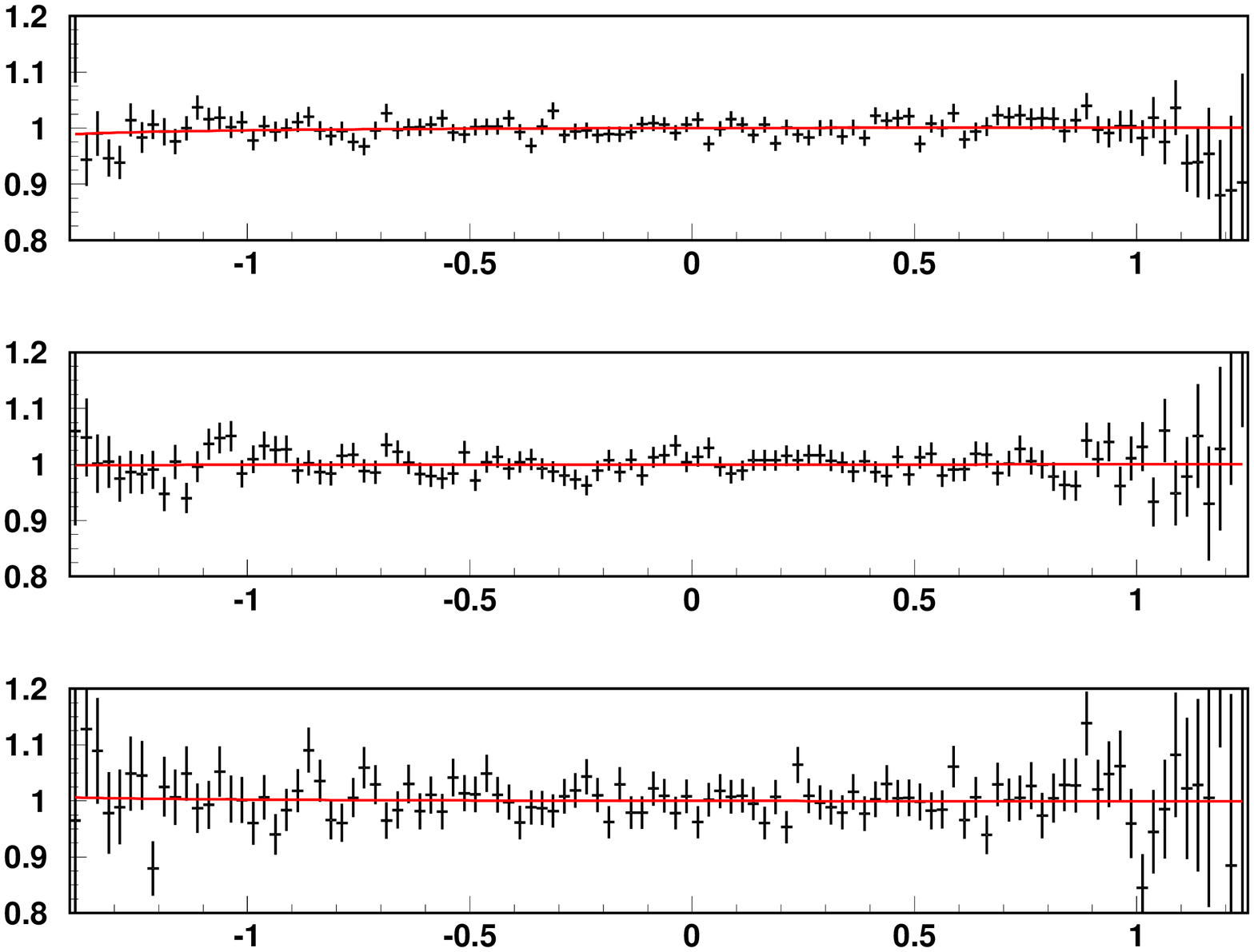}}
\put(-38,18){$u$} \put(-38,105){$u$} \put(-38,193){$u$}
\put(-325,75){Super-sample III: $\chi^2/n.d.f.=$ 85/104}
\put(-325,162){Super-sample II: $\chi^2/n.d.f.=$ 88/104}
\put(-325,248){Super-sample I: $\chi^2/n.d.f.=$ 97/104}
\put(-385,85){$R_4(u)$} \put(-385,170){$R_4(u)$}
\put(-385,260){$R_4(u)$}
\end{center}
\vspace{-10mm} \caption{Fits of the normalized quadruple ratios for each
super-sample.} \label{fits}
\end{figure}

\begin{figure}[t]
\begin{center}
\resizebox{.9\textwidth}{!}{\includegraphics{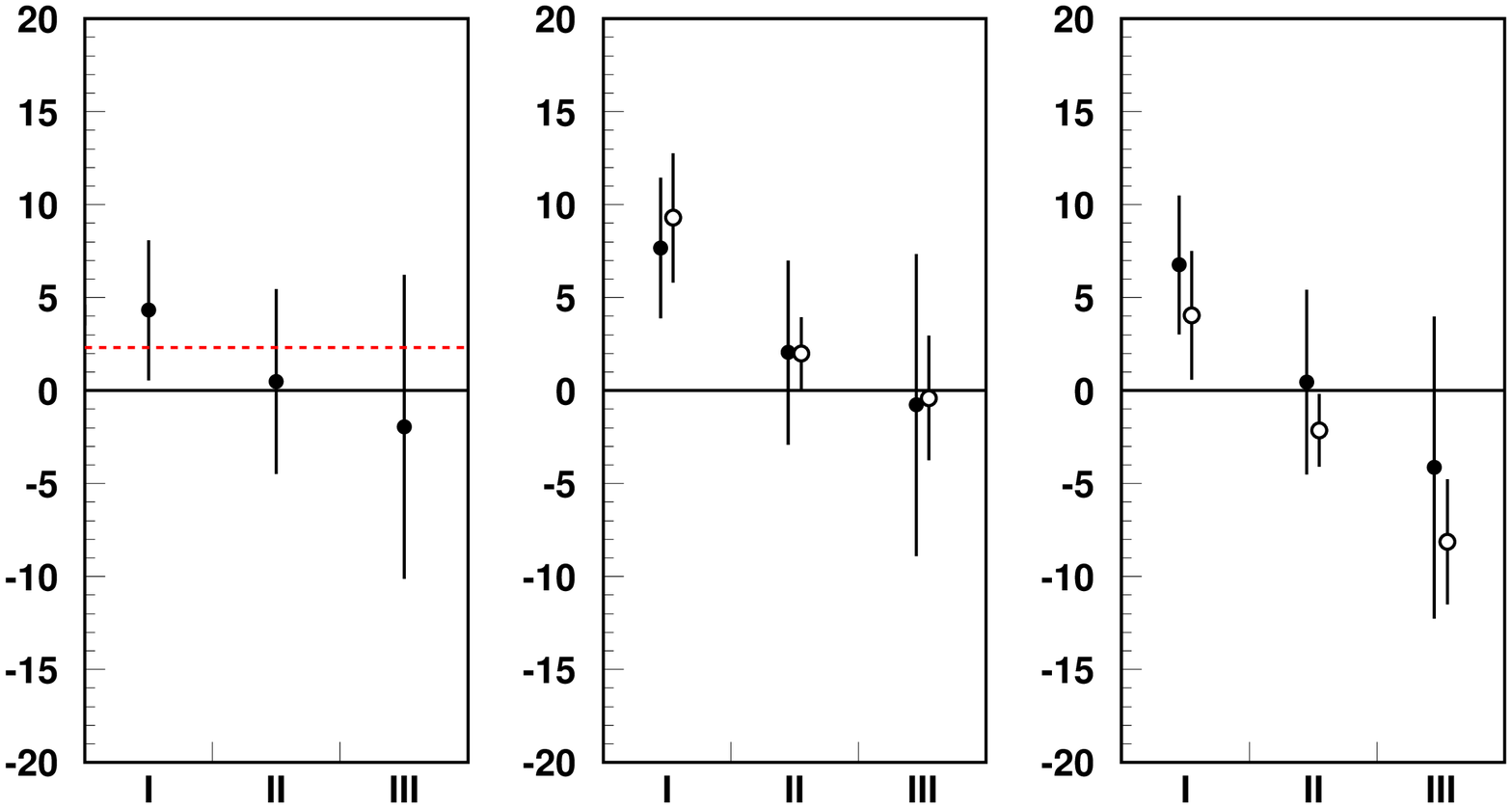}}
\put(-112,22){I\hspace{0.85cm}II\hspace{0.7cm}III}
\put(-238,22){I\hspace{0.85cm}II\hspace{0.7cm}III}
\put(-363,22){I\hspace{0.85cm}II\hspace{0.7cm}III}
\put(-103,40){Super-sample}
\put(-85,0){(c)}\put(-212,0){(b)}\put(-338,0){(a)}
\put(-140,225){$\Delta_{UD}, 10^{-4}$} \put(-267,225){$\Delta_{SJ},
10^{-4}$}\put(-393,225){$\Delta g, 10^{-4}$}
\end{center}
\vspace{-6mm} \caption{(a) Slope difference $\Delta g$ for each
super-sample; (b) left-right instrumental asymmetry and (c) up-down
instrumental asymmetry for experimental data (full circles) and Monte
Carlo events (open circles) (these instrumental
asymmetries cancel in the quadruple ratio (6)).}
\label{dg}
\end{figure}

Since each super-sample listed in Table~\ref{t1} allows an independent
asymmetry measurement to be made
according to the full procedure described above,
$\Delta g$ is obtained
for each of these. The resulting fits for these data sets are shown
in Fig.~\ref{fits}. The corresponding best fit values for $\Delta g$
are presented in Table~\ref{t2} with the statistical error only and
are plotted in Fig.~\ref{dg} (a).
\begin{table}[h]
\begin{center}
\begin{tabular}{|c|c|}
\hline
Super-sample & $\Delta g\cdot 10^4$ \\
\hline \hline
I & $4.3\pm 3.8$ \\
\hline
II & $0.5\pm 5.0$ \\
\hline
III & $-2.0\pm 8.2$ \\
\hline \hline
Total & $2.3\pm 2.8$ \\
$\chi^2$/ndf: 0.7/2 & \\
\hline
\end{tabular}
\caption{Slope difference, $\Delta g$ with its statistical
error for each
super-sample and the weighted average.} \label{t2}
\end{center}
\end{table}

The final result is obtained as the weighted average of the $\Delta
g$ values for the three super-samples:
\begin{equation}
\Delta g=(2.3\pm 2.8)\cdot 10^{-4}.
\end{equation}
where the quoted error is only statistical.

Since the two $\pi^0$'s are indistinguishable, a linear term in $v$
cannot appear in the matrix element (1), and no asymmetry is
expected for the linear $|v|$ slopes of $K^+$ and $K^-$ decays. A
measurement of the difference in these slopes should be considered,
therefore, as a check of the cancellation of spurious charge
asymmetries. A quadruple ratio of $|v|$ spectra is constructed and
fitted with a linear function. The difference in slopes, averaged
over the three super-samples, is found to be
$(1.1\pm4.7)\cdot10^{-4}$, which is consistent with zero.

As an additional cross-check of the method, quantities sensitive to
instrumental left-right and up-down charge asymmetries, which cancel
in the quadruple ratio (7), were evaluated. The asymmetry $\Delta_S$
between $K^+$ and $K^-$ deflected towards the Sal\`{e}ve mountain
(no matter which way the beams travel in the achromats) can be
extracted from a fit of the double ratio $R_{US}\cdot R_{DS}$ to the
function $R_S[1+\Delta_S\cdot~u/(1+gu+hu^2)]^2$, where $R_S$ is a
normalization parameter. $\Delta_S$ describes a combination of the physical
CP-violating asymmetry between $K^+$ and $K^-$ and the asymmetry of
the apparatus response to particle of both charge signs deflected
towards the Sal\`{e}ve. An analogous asymmetry $\Delta_J$ can be
defined for $K^+$ and $K^-$ deflected towards the Jura mountains by
fitting the double ratio $R_{UJ}\cdot R_{DJ}$. Then in the
half-difference $\Delta_{SJ}=(\Delta_S-\Delta_J)/2$ any physical
CP-violating asymmetry cancels, while the left-right instrumental
asymmetry remains. Similarly, an up-down beam line asymmetry can be
defined as $\Delta_{UD}=(\Delta_U-\Delta_D)/2$, where $\Delta_U$ and
$\Delta_D$ are extracted from a fit to the double ratios
$R_{US}\cdot R_{UJ}$ and $R_{DS}\cdot R_{DJ}$, respectively. These
instrumental asymmetries are small, and are quantitatively well
reproduced by the Monte Carlo simulation, as shown in Fig.~\ref{dg}
(b) and (c).

\section{Systematic uncertainties}

The measured asymmetry (8) should be free from systematic biases thanks
to the cancellation of instrumental asymmetries implemented in the
detector design and data analysis. Nevertheless, a number of
quantitative checks of
possible systematic contributions to the measured $\Delta g$ value
have been performed.

Since only information on photon clusters is used for the $u$
calculation, systematic uncertainties associated with
the LKr response have been considered:
\begin{itemize}
\item Changes of the measured asymmetry are studied by varying the size of
the $u$ bins with respect to their nominal value of 0.025,
including the use of bin sizes proportional to the $u$ resolution (which
vanishes at the lower $u$ edge and reaches $\pm 0.04$ at high $u$).
The maximum observed difference, $\delta\Delta g=0.4\cdot 10^{-4}$,
is taken as an upper limit for the systematic uncertainties
associated with the $u$ calculation.
\item An uncertainty $\delta\Delta g=0.1\cdot 10^{-4}$ arises from the
correction applied to the measured photon energies to account for
uncertainties in the LKr non-linear response at low photon
energies (typically $2\%$ at 3 GeV and becoming negligible
above 10 GeV).
\item By varying the cut on the minimum allowed distance between
LKr clusters the measured asymmetry is found to vary by less than
$\delta\Delta g=0.5\cdot 10^{-4}$.
\item The effect of wrong photon pairings in the reconstruction
of the $\pi^{0}-\pi^0$ pair
is found to be negligible ($\delta\Delta g<0.1\cdot10^{-4}$) from
the study of a large sample of simulated events.
\item The measured asymmetry is found to be insensitive to changes
of the radial cut around the LKr central hole.
\end{itemize}

Effects related to the magnetic spectrometer do not affect the
result directly, since the charged track is only used for tagging,
and not for the $u$ calculation. Upper limits from effects such as
spectrometer alignment and momentum scale are found to be
negligible ($\delta\Delta g<0.1\cdot10^{-4}$). The
total systematic uncertainty associated with the charged track geometrical
acceptance and with beam geometry is found to be
$\delta\Delta g<0.3\cdot10^{-4}$ by varying the radial cuts which
define the acceptance around the beam tube. An
estimate of a possible bias
from the contamination of $\pi\rightarrow\mu\nu$ decays is
obtained from a Monte Carlo simulation and does not exceed $\delta\Delta
g=0.5\cdot 10^{-4}$. An upper limit $\delta\Delta g=0.1\cdot
10^{-4}$ is obtained from the uncertainties on the
permanent magnetic fields in the $K^{\pm}$ decay region
by artificially varying their map in the event reconstruction. In
addition, it is checked that the result is stable with respect to
wide variations of the accepted $\pi^\pm\pi^0\pi^0$ invariant
mass interval.

Systematic uncertainties associated with the presence of
accidental tracks and LKr
clusters are found to be $\delta\Delta g=0.2\cdot 10^{-4}$ by varying
the allowed number of additional particles in the event and by checking
the stability of the result with different timing cuts.

The trigger inefficiency, if it depends on $u$ and on the pion charge
and if it is not stable in time, could bias
the measured asymmetry. Each component of the
trigger logic is studied separately for possible systematic biases:
\begin{itemize}
\item The inefficiency ($\sim 0.25\%$) to charged tracks of the level 1 trigger
is measured for each sub-sample as a function of the $x$ and $y$ coordinates
of the charged track impact point on the hodoscope plane for each sub-sample.
This is done by studying all one-track events in control
samples recorded by triggers which
do not use the charged hodoscope. Any effect of this inefficiency
on $\Delta g$ is corrected by reweighting
the $K^{\pm}\rightarrow\pi^{\pm}\pi^0\pi^0$ events and the result is found
to change by $\delta\Delta g=(0.1\pm0.1)\cdot 10^{-4}$ from the
nominal value. A similar result is obtained from the Monte Carlo simulation.
\item The largest systematic uncertainty $(\delta\Delta g=1.3\cdot 10^{-4})$
in the present analysis is conservatively estimated as an upper limit
due to any unknown effect associated with the inefficiency of the
level 1 neutral trigger based on a requirement on the number of LKr
clusters. This inefficiency is found to vary from
0.7$\%$ at the beginning of the data-taking period, to $3\%$ at the
end of the run, as measured using a control sample. This estimation
is limited by the statistics of the control sample and further
analysis is expected to reduce this error.
\item The average level 2 inefficiency is $\sim5.7\%$.
The main part ($\sim70\%$ of it)
is due to local inefficiencies in the DCHs, which could affect the
asymmetry measurement if they are not stable in time. The DCH wire
inefficiency is properly simulated by the Monte Carlo and a shift
$(0.1\pm0.2)\cdot10^{-4}$ is measured between the result obtained from
the total sample of Monte Carlo events and the sample containing only events
satisfying the level 2 trigger. The total uncertainty from the level 2
inefficiency, including other possible sources (timing effects
between detectors, data buffer overflows, inefficiency of the level 2
algorithm) is found to be less than $\delta\Delta g=0.4\cdot10^{-4}$.
\end{itemize}

\begin{table}[t]
\begin{center}
\begin{tabular}{|l|c|}
\hline LKr related effects & 0.7 \\
\hline Beam geometry and charged track acceptance & 0.3 \\
\hline $\pi\rightarrow\mu$ decay & 0.5 \\
\hline Accidentals & 0.2 \\
\hline Trigger Level 1 & 1.3 \\
\hline Trigger Level 2 & 0.4 \\
\hline \hline Total systematics uncertainty & 1.6 \\
\hline External uncertainty & 0.3 \\
\hline
\end{tabular}
\caption{Systematic uncertainties to the measured $\Delta g$ value (in
units of 10$^{-4}$).} \label{t3}
\end{center}
\end{table}

All systematic uncertainties are summarized in Table~\ref{t3}. The
overall uncertainty, obtained by summing all contributions in
quadrature, is conservatively taken to be $1.6\cdot 10^{-4}$. An
additional external uncertainty of $0.3\cdot 10^{-4}$ arises from
the known precision on $g$ and $h$~\cite{13}\footnote{By using the
measured value~\cite{15} $g=0.645\pm0.010$, which is consistent with
the PDG value~\cite{13}, this external error becomes negligible.}.

\section{Conclusion}

By studying  $4.714\cdot10^7$
$K^{\pm}\rightarrow\pi^\pm\pi^0\pi^0$ decays the difference between
the slope parameters in the Dalitz plots of $K^+$ and $K^-$ decays has been
measured:
\begin{equation}
\Delta g=(2.3\pm 2.8_{stat.}\pm 1.3_{trig.(stat.)}\pm 1.0_{syst.}
\pm 0.3_{ext.})\cdot 10^{-4}.
\end{equation}
The corresponding asymmetry parameter $A_g$, which describes a
possible direct CP-violation in these decays, is found to
be\footnote{A preliminary result reported earlier~\cite{16}
neglected the $u^2$ term in the fitting function (6).}:

\begin{equation}
A_g=(1.8\pm 2.2_{stat.}\pm 1.0_{trig.(stat.)}\pm 0.8_{syst.}\pm
0.2_{ext.})\cdot 10^{-4}=(1.8\pm2.6)\cdot10^{-4}.
\end{equation}
This result is almost one order of magnitude more precise than
previous measurements and is consistent with the predictions of the
Standard Model.
\subsection*{Acknowledgments}
We gratefully acknowledge the CERN SPS accelerator and beam line
staff for the excellent performance of the beam. We thank the
technical staff of the participating laboratories and universities
for their effort in the maintenance and operation of the detectors,
and in data processing.

\end{document}